\documentclass[a4paper]{article}

\usepackage[english]{babel}
\usepackage[utf8]{inputenc}
\usepackage{amsmath}
\usepackage{graphicx}
\usepackage[colorinlistoftodos]{todonotes}
\usepackage{hyperref}
\usepackage{fullpage}
\usepackage{parskip}
 \usepackage{float} 
 \usepackage{graphicx}
\usepackage{multirow}
\usepackage{tabularx}
\usepackage{soul}
\usepackage{rotating}
\usepackage{cite}

\title{The 2021 RecSys Challenge Dataset: \\Fairness is not optional}

\author{Luca Belli, \and 
    Alykhan Tejani\thanks{Authors contributed to the research equally},  \and 
    Frank Portman\footnotemark[1],  \and 
    Alexandre Lung-Yut-Fong\footnotemark[1] ,  \and 
    Ben Chamberlain,  \and 
    Yuanpu Xie,
     \and Kristian Lum,  \and 
     Jonathan Hunt, \and 
     Michael Bronstein,  \and 
     Vito Walter Anelli, \and 
     Saikishore Kalloori,  \and 
     Bruce Ferwerda, \and 
     Wenzhe Shi}
\date{}

\begin{document}
\maketitle

\begin{abstract}
After the success the RecSys 2020 Challenge, we are describing a novel and bigger dataset that was released in conjunction with the ACM RecSys Challenge 2021. This year's dataset is not only bigger (\textasciitilde 1B data points, a 5 fold increase), but for the first time it take into consideration fairness aspects of the challenge.
Unlike many static datsets, a lot of effort went into making sure that the dataset was synced with the Twitter platform: if a user deleted their content, the same content would be promptly removed from the dataset too. In this paper, we introduce the dataset and challenge, highlighting some of the issues that arise when creating recommender systems at Twitter scale. 

\end{abstract}

\section{Intro}

Recommender systems are a central component of  modern social media platforms. On Twitter, when a user logs in, they are presented with a series of Tweets (the Home Timeline). Users can decide if the content they are shown is sorted in reverse chronological order or is algorithmically ranked by a recommender system. If they opt for the latter, content is presented in decreasing (predicted) order of value for the user: the Tweets that the recommender system predict are more relevant for the reader are put in a higher position. 

For users who select algorithmically ranked Home Timelines, the quality of the recommender system's rankings directly impacts the quality of their experience on the platform. From the point of view of the person who is receiving the Tweets, a well-performing recommender system may result in a more interesting or enjoyable experience on the platform. It may also offer opportunities to participate conversations they would like to contribute to or access to important news, perspectives, or connections to new communities they would like to be a part of. From the point of view of the Tweet authors, the recommender system impacts the extent to which their Tweets reach a receptive audience. A better job reaching that audience can lead to more social influence.  

Given the impacts of recommender systems in social media, it is important that they work well for all users, not just some subset. That is, it is important that recommender systems behave ``fairly" by some definition of the word. Recent years have seen a flurry of methodological work in ``fair ML.", e.g. \cite{CorbettDavies2018TheMA}. Though there has been high quality work developing notions of fairness for recommender systems (see, e.g., \cite{singh2018fairness, biega2018equity, sapiezynski2019quantifying, asudeh2019designing}), fairness in classification tasks has received the bulk of the attention. The existence of a publicly available dataset with a pre-defined ``fairness task"-- the COMPAS dataset-- undoubtedly played no small part in spurring research into fair classification. In hopes of providing a similar resource to researchers interested in developing state-of-the-art methods for fairness in recommender systems at real scale, in this paper we introduce a fairness challenge and accompanying dataset comprised of Twitter data.

In Section \ref{problem-definition} we formally introduce the task of the  challenge, explaining the differences between the challenge and Twitter's product usage.
%In Section \ref{challenge} we explore the challenges that come with building recommender systems at Twitter scale. 
In Section \ref{challenge-dataset} describes the challenge's dataset. Finally in Section \ref{challenge} we discuss the 2021 RecSys Challenge in detail, providing details on how submissions are evaluated and the baseline that we provided.
\section{Problem Definition} \label{problem-definition}
%Here we talk about the problem of ranking
%We should also say that is is an incomplete ranking (e.g. not all the candidates are there)
%We should talk about feedback loops too, because the current candidates are output of previous models

Twitter is what's happening. When people log in on Twitter, they are presented an updating display of Tweets— content produced by other users. Twitter users indicate the content they would like to see in multiple ways: users can follow other users, they can follow list of multiple users, or they can follow specific Topics (e.g. ``Sports'', ``Art''). Tweets from the users or topics they follow can appear on their timeline.

Users can decide if this timeline is sorted by reverse chronological order or if it is algorithmically ranked. If they opt for the former, their timeline consists of all the latest Tweets, arranged so that the newest appears first. If they opt for the latter, content is presented in decreasing (predicted) order of value for the user: the Tweets that our models predict are more relevant for the reader are  put in a higher position. Given the fast pace of Twitter and the speed with which the conversation changes, it is very important that the most relevant and timely content is presented to the reader. %This might be at odds with the desire odds the authors who are competing to have their content shown.

%In practical terms, this means that each time a user requests a new timeline, a score (say between 0 and 1) are assigned to Tweets candidates and they are sorted according to it.
%For simplicity, given a user $u$ and a timestamp $t$, we define the candidates $C_t(u)$ as the Tweets that have been produced by the users that $u$ follows since the last time that $u$ refreshed their timeline. In reality the situation is more complex, as the users are shown with a wider variety of Tweets (e.g. Tweets from Topics they follow). For the ease of notation, we are going to drop the dependence on $t$. 

%Therefore at serving time, the ranking model $R$ has access to the full set of $C(u)$.
%The problem of ranking relevant content  is very significant for Twitter: that is why for the RecSys 2021 Challenge we pose the following problem: 
For the purposes of the RecSys 2021 Challenge, we define the core problem for this challenge as follows: 

\emph{Given a (reader, Tweet) pair, ``fairly" predict the probability that the reader will engage with the Tweet.}

We define engagement as one of four different types: \textit{Like} (reader clicks the Like button), \textit{Reply} (reader replies to the Tweet, thus creating a new Tweet themselves), \textit{Retweet} (reader shares the original Tweet with their followers) and \textit{Quote} (reader adds their comment or media before sharing the original Tweet). We define what we mean by ``fairly" in section \ref{challenge}.

\section{The Data} \label{challenge-dataset}

\begin{figure}[h]
    \centering
    \includegraphics[width=0.8\textwidth]{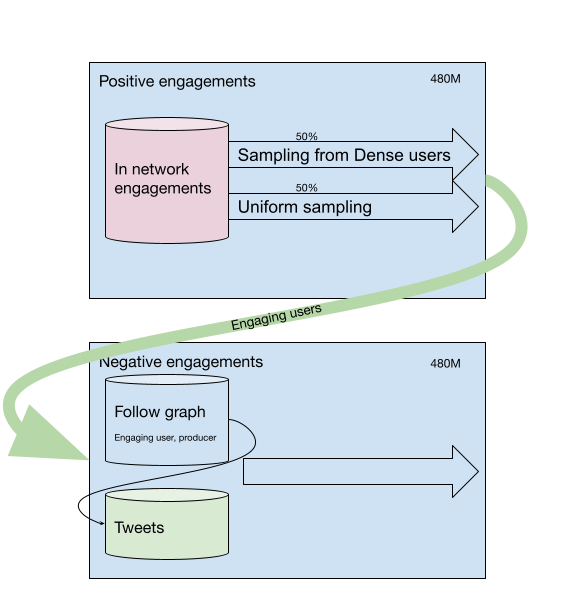}
    \caption{Data generation sampling pipeline}
    \label{fig:samplingpipeline}
\end{figure}

The data for the challenge represents a snapshot of a 4 week period on public users on Twitter. The first 3 weeks were used for the training set and the last week was used for testing and validation. We were aiming for approximately 1 billion records, equally balanced between positives and (pseudo)negatives. %In contrast to last year's challenge \cite{belli2020privacyaware}, as described below, we also ensured that approximately half of the positive samples came from so-called ``dense users". These are users who had at least 10 incoming or outgoing engagements with other users during the time period. 
Figure \ref{fig:samplingpipeline} describes the data generation process for the challenge. 

%\subsubsection{Sampling}

\subsection{Features} \label{dataset:features}
Features available for prediction are divided into three separate feature groups: \textit{user-}, \textit{Tweet-} and  \textit{engagement features}. There are two instantiations of \textit{user features}, one for the author (producer) and one for the reader (consumer) of the Tweet. \textit{Tweet features} groups all the attributes describing the original Tweet, that is possibly engaged with by the consumer.
Tweet tokens represent the text of the Tweet. We wanted to find a reasonable trade-off between obfuscation (i.e. including plain text would have made re-identification trivial) and making sure that the content of the Tweet would be available for NLP tasks. As in the previous year, we released the text in form of the BERT tokens: after tokenizing the text, we release the list of the BERT IDs for each Tweet.

Finally, the \textit{engagement features} contain all the details of the engagement itself. These are the outcomes to be predicted. The dataset features are described in detail in Table~\ref{tab:feature_list}.

\begin{table*}
\resizebox{\textwidth}{!}{%
\begin{tabular}{|c| l | l | l |}
\hline
%\multicolumn{1}{|>{\Large}c|}{\textbf{Feature } } & %\multicolumn{1}{>{\Large}c|}{\textbf{Name} } & %\multicolumn{1}{>{\Large}c|}{\textbf{Signature} } & %\multicolumn{1}{>{\Large}c|}{\textbf{Description} } \\
\textbf{Feature } & \textbf{Name} &\textbf{Signature} &\textbf{Description} \\
 \hline
{\multirow{5}{*}{\textbf{User}}} & userId & \textit{string}   & User identifier (hashed) \\
                                     & follower count & \textit{int}                  & Number of followers of the user                                  \\
                                      & following count & \textit{int}                  & Number of accounts this  user is following                        \\
                                       & is verified & \textit{bool}                      & Is the account verified?                                         \\
                                    &  account creation time & \textit{int}  & timestamp of the creation time of the account \\ 
  \hline
{\multirow{7}{*}{\textbf{Tweet}}} &  tweetId & \textit{string}                            & Tweet identifier (hashed)                                                 \\
                                     &  presentMedia & \textit{list[string]}                  & Tab-separated list of media types; media type can be in (Photo, Video, Gif)                                  \\
                                      &  presentLinks & \textit{list[string]}                  & Tab-separated list of links included in the tweet (hashed)                       \\
                                          &  presentDomains & \textit{list[string]}                  & Tab-separated list of domains (e.g. twitter.com) included in the tweet (hashed)   
                           \\           &  tweetType & \textit{string}                     & Tweet type, can be either Retweet, Quote, Reply, or Toplevel                                         \\
                                       & language & \textit{string}  & Identifier corresponding to inferred language of the tweet \\
                                       &  tweet timestamp& \textit{int}                      & Unix timestamp, in seconds of the creation time of the Tweet                                         \\
                                       &  tweet tokens & \textit{list[int]}   & Ordered list of Bert ids corresponding to Bert tokenization of Tweet text \\
                                            &  tweet hashtags & \textit{list[string]}   & Tab-separated list of hashtags present in the tweet \\
\hline                                      
\multirow{5}{*}{\textbf{Engagement}} & reply engagement timestamp & \textit{int}  & timestamp of the Reply engagement if one exists  \\
 & retweet engagement timestamp & \textit{int}  & timestamp of the Retweet engagement if one exists                                  \\
 & quote engagement timestamp & \textit{int}  & timestamp of the Quote engagement if one exists                     \\
  & like engagement timestamp & \textit{int}  & timestamp of the Like engagement if one exists     
  \\
  
  & engageeFollowsEngager    & \textit{bool}                  & Does the account of the engaged tweet author follow the account that                                         \\
                                      &  &                     & has made the engagement?                                         \\ \hline
\end{tabular}
}
\caption{List of features provided for the challenge dataset}
\label{tab:feature_list}
\vspace{-2em}
\end{table*}

\subsection{Sampling} \label{sampling}

%\textbf{Pseudonegative features} 
To create a useful dataset for a supervised learning task, we need a mixture of both positive and negative samples. On Twitter, engagements for public profiles are public so the positive case is straightforward. On the other hand, the converse is not true. In this case a negative sample is a \emph{(reader, Tweet)} that was seen, but not engaged with. This information is not publicly available: authors and other browsers of Twitter do not know which users \emph{saw} which Tweet. Therefore releasing  ``true'' negatives would have constituted releasing nonpublic information about our users.

For the negative features we did the following: for each reader, we collected the public Tweets that were produced by their followees during the sampling period. Some of those might have been (publicly) engaged with, so we removed them from this pool (they may separately have been in the positive sample). The rest of Tweets were Tweets that were not engaged with, but here is the catch: \emph{this group contains both Tweets that were seen and not seen by the reader}. Negative examples were sampled from this group. We call the collection of negatives ``pseudonegatives", since they represent tweets that were not engaged with, though the user may or may not have had the opportunity to engage with them. 

It is very likely that a model trained on this dataset would under-perform with an equivalent model trained on a set of pure negative samples. Since the goal of this challenge is not to implement the solutions in the production case, the loss of predictive performance is easily worth not revealing private usr information. Note that special care was taken to make sure the readers in the negative set were the same users that were in the positive set, to avoid the cold start problem. 

%Figure \ref{fig:samplingpipeline} describes the data generation process for the challenge. 

\subsection{Data set release conditions}

%\textbf{Developer Agreement andPolicy} 
Before having access to the dataset, participants are required to be approved for the use of Twitter's API, and thus adhere to the Developer Agreement and Policy which includes (but is not limited to) the restrictions to “off-Twitter matching” to data that has been directly provided by the person and/or public data. This policy serves to prevent misuse of the dataset, such as inference attacks on private datasets.\footnote{https://developer.twitter.com/en/developer-terms/agreement}

%\textbf{Scrubbing deleted content} 
Over time, a user might decide to delete one or more of their Tweets, or decide to delete their profile (or make it private). Again, since we want the dataset to only contain public data at all times, the dataset itself is constantly updated to track what is on the Twitter platform. Specifically this means that the dataset is shrinking as we are not adding new content over time, only deleting data that is not available anymore. Furthermore participants are required to keep their dataset up-to-date as required in the Developer Agreement and Policy and we provide a file that includes only details on the content that needs to be scrubbed.

This of course means that people accessing the dataset at different time will have access to different datasets. Specifically people who trained on the first day might have a dataset advantage with respect to people who trained on the last available day. Given the magnitude of the dataset, and the tendency of such models to be constantly re-trained over the course of the challenge, we do not expect this to impact things too much as participants are always required to use the latest version available. 

\subsection{Comparison to data from Twitter's Recsys2020 challenge}
%We now describe how the dataset and the challenge were formulated to address the aforementioned issues. 
This challenge has much in common with last year's: Recsys Challenge 2021. For those familiar with Recsys Challenge 2020 \cite{belli2020privacyaware}, briefly summarize the major similarities and differences from this year's version in Table~\ref{tab:differences}.

\begin{table}[h!]  \label{tab:differences}
\begin{tabularx}{\textwidth}{|l|X|X|}
\hline
& \textbf{2020 Challenge} & \textbf{2021 Challenge} \\
\hline
size & $\sim$ 200 million records & $\sim$ 1 billion records \\
\hline
Process &         Scrubbing, pseudonegatives, pre-featurized text tokens & Scrubbing, pseudonegatives, pre-featured text tokens \\
\hline
Sampling &         Fully uniform sampling 
            & Half uniform sampling and  half of records sampled from a denser subgraph \\
\hline
Submission &        Submit TSV file with predictions (no latency constraint) & Submit model that  must run within 24h on fixed hardware \\
\hline
Metrics &        Accuracy metrics determine winner (PR-AUC and RCE) & Accuracy metrics coupled with a popularity based fairness metric determine winner \\
\hline
\end{tabularx}
\caption{2020 Challenge versus 2021 Challenge}
\end{table}

%
% While the challenge is meant to mimic the real-world case of Twitter  as much as possible , there are some important differences and simplifications that we were necessary to build a dataset for this challenge. 

%\subsubsection{Features}
 \subsection{Differences with Respect to the Twitter Case} \label{differences-twitter}
 This dataset captures the main characteristics of those used at Twitter to build recommender systems. However, because this data was compiled for external users, some simplifications were necessary.
 
 To respect users' privacy, only (a strict subset of) features that are already public (e.g number of likes for a given Tweet) are made available in the dataset. Machine learning practitioners within Twitter have access to a larger number of features, both public and private. %For a list and a description of features in the dataset, refer to Section \ref{dataset:features}. 

In the interest of privacy,  we obfuscated all the field that would make the re-identification trivial (e.g. the Tweet id) before releasing the dataset. We are aware that this approach is ineffective against linkage attacks (e.g. \cite{Narayanan2006HowTB} and \cite{Sweeney1997GuaranteeingAW}) and the data can be easily reconstructed. However there are reduced privacy concerns since all the data that we started with is already public. Contrast this dataset with a dataset where a mixture of both private and public data was released (e.g. Netflix challenge) and public features were used to infer private ones. 

This dataset is also much smaller scale than data availability to machine learning practitioners within Twitter. Even with its almost 1B rows and more than 200 GB, it only represent a small fraction of the all the content that is produced on Twitter at any given moment. 
%\subsubsection{Scale} 
That said, to the best of our knowledge — at the time of writing — this dataset is the biggest publicly available social media-based dataset for reccomender systems. % We will focus our attention on the challenges that come with scale later in Section \ref{challenges-at-scale}. 
%\subsubsection{Incomplete candidates}

When a production model ranks the Tweets for a user $u$, it has access to the full set of candidates (e.g. all Tweets that were produced by $u$'s followees since last time they requested a fresh timeline). In the dataset we cannot guarantee that all the candidates are present — both at training and validation/test phases. %For more details on how the dataset is sampled, refer to Section \ref{sampling}.
%\subsubsection{Other type of content}

Finally, for the purpose of this challenge, we are only including content from users that are explicitly followed by the reader. In reality the situation is more complicated, and other kind of content might be presented to the user. For example, as mentioned above users may be presented with Tweets that belong to Topics they follow. A user's timeline may also include Tweets that someone they follow liked or responded to, even if they do not follow that person themselves. Additionally, paid advertisements may also appear among the set of Tweets displayed to a user. 
\section{The RecSys 2021 Challenge} \label{challenge}

In the past, ML challenges focused on optimizing for a single metric (e.g. accuracy) without considering other factors. For the first time in the history of the RecSys challenge, we have included fairness considerations into the evaluation metrics. To the best of our knowledge, this is the largest challenge to date that has incorporated a fairness component. 

Fairness is a societal concept, not an optimization one \cite{10.1145/3287560.3287598} and it can come in many different shape or forms. 
When defining what fairness means for the purpose of this challenge, we considered three criteria:

\begin{itemize}
    \item We wanted to ease participants into the fairness space, without requiring too much prior knowledge or special infrastructure.
    \item At the same time, we wanted the problem to be meaningful and have a real significance for Twitter. In short, we did not want to focus on a ``toy problem.''
    \item Finally we wanted to rely on the available features (see Section \ref{dataset:features}). This prevented us from considering any metrics that depend on demographic data, as those features are not included in the datset.  
\end{itemize}

A popularity-based (measured as the number of followers for an author) metric, has all the above characteristics. In this scenario, the quality of the recommendations should be independent from the popularity of the authors. Said in another way, users should not be getting worse recommendations for being less popular on the platform.

Concretely, we divide users into 5 quantiles (based on the number of followers of the authors in the test set). In the parlance of fair ML, the ``sensitive attribute" for the purposes of evaluation is the binned user popularity variable.  The number of rows in the datset is not be equal for each cohort since more popular authors have a larger audience which typically translates into more opportunity for incoming engagement.

We also want to acknowledge that having more followers might be itself an effect of a feedback loop from the recommender system: more popular authors might be shared more thus gaining more even popularity. 
We do not want to suggest that the producer popularity constraint is the only, or even the most important, aspect to consider for serving fair recommendations. At the same time we do feel that it allows us to make an important step forward with the recommender system community about the recommendations not being only about accuracy.

\subsection{Metrics} \label{metrics}
For last year's challenge, we used Relative Cross Entropy (RCE) and Area under the Precision Recall Curve (PR-AUC) for each engagement,  ranked these and then summed the positions to get the final score. For more details on how those metrics were calculated, refer to \cite{belli2020privacyaware}.

However to calculate PR-AUC we used the Python \texttt{scipy} library \cite{2020SciPy-NMeth}. The method implemented there has a drawback that constant predictions you will get a PR-AUC of 0.5. This was exploited by a lot of teams who just submitted constant predictions - meaning they would have very bad RCE but a PR-AUC of 0.5.

To prevent the same problem this year, we used average precision and RCE. 

\subsubsection{Relative Cross Entropy}
Relative Cross Entropy (RCE) corresponds to the improvement of a prediction relative to the straw man, or the naive prediction, measured in cross entropy (CE). The naive prediction corresponds to the case that does not take into account the user and Tweet features, e.g., it always predicts the average (observed) CTR of the training set. Suppose the average CE of the naive prediction is $CE_{naive}$ and average CE of the prediction to be evaluated is $CE_{pred}$, then RCE is defined as $(CE_{naive} - CE_{pred}) \times 100/CE_{naive}$. Note that the lower the CE the better the quality of the predictions, so the higher the RCE. The benefit of using RCE is that we can obtain a confident estimate of whether the model is under or over performing the naive prediction.

\subsubsection{Average Precision}
We used  Average Precision from the \texttt{scikit-learn} package \cite{scikit-learn}, that ``summarizes a precision-recall curve as the weighted mean of precisions achieved at each threshold, with the increase in recall from the previous threshold used as the weight." While this metric is basically the same as PR-AUC, it helps with the aforementioned problem with constant predictions.

\subsubsection{Final score}
As discussed, we wanted the performance of the recommendations to be independent of the popularity of the user. %of the author of the Tweet.
After dividing the authors in five quantiles (based on the number of followers of the authors on the test set), average precision and RCE are calculated for each group. The two scores are then averaged across groups (so that entries that have great performance on  popular users but perform poorly on smaller accounts are penalized) and ranked. The final score is the sum of the position according to each score. So for example, if someone ranked first in the RCE ladder and third in the AP one, their final score would be 4.

\subsection{Baseline}

In this section, we will describe a simple baseline model architecture that we use for this competition. We used a similar baseline as last year \cite{belli2020privacyaware}. Specifically, it mainly consists of the feature processing and prediction components.

\subsubsection{Feature processing}
We apply different feature processing methods based on different feature types. 
For Numeric Features (e.g. follower count), we apply $z$-score normalization first to shift the feature into a reasonable range, and then transform each normalized feature with a tiny neural network, which consists of one hidden layer of size 2 and $tanh$ activation. The missing features are populated with 0.

Categorical features (e.g., Tweet type, language) and binary features are represented using one-hot encoding. Please note that we reserve a bucket for any missing categorical and binary feature values.  ID features are hashed into a given number of buckets due to its extremely high cardinality. Tweet text feature are tokenized and embedded using a pre-trained BERT model ~\cite{devlin-etal-2019-bert}. 

\subsubsection{Model}
The one-hot encoded representations of the aforementioned  categorical, binary and ID features are concatenated and converted to dense representation of size 16, which are then concatenated with the normalized numeric features. We omit the text encoding module in the submission for simplicity and computational efficiency. The obtained processed feature vector is then fed into a 3 layer multi-layer perceptron, which consist of 3 dense layers of size 128, 64, and 32, respectively. The activation functions are leaky ReLU.  The final prediction of the model is of size 4, corresponding to 4 types of engagement (Retweet, Reply, Like and Retweet with comment). In the final prediction layer, the activation function is chosen as sigmoid rather than softmax since different types of engagements are not mutually exclusive to each other.  The model is trained with Adam optimizer ~\cite{adam} using Huber loss for 1 million steps, and the learning rate is 0.001.

\subsection{Latency constrains}
In the RecSys Challenge 2020 \cite{belli2020privacyaware}, after the training and validation phase, participants were asked to upload their prediction on the test set. This gave total freedom to participants to train and test their model on their own hardware and did not impose any restrictions. In practice,  Twitter's models need to be able to score content on the order of milliseconds. While we did not want to impose the same production constraints, we also wanted to nudge participants away from slower and more complex models (e.g. ensemble ones) even if this comes at the cost of accuracy.

While during the training stage, participants were free to train on hardware of their choosing, for the testing phase, we asked people to upload their code, which would be run on a dedicated Docker instance. Furthermore, we imposed a 24h timeout limit for the entirety of the dataset. The dedicated instance had 1 CPU with  64Gb of RAM. 

%'docker', 'run',
%  '--network', 'none',
%  '--memory', '{memory_limit}g',
%  '--cpus', '{num_cpus}',
%  '--memory-swap', '{memory_limit}g',
%  '--shm-size', '1024M',
%  '--name', 'name_{submission_identifier}',
%  '-v', '{work_dir}:/home/{submission_identifier}',
%  '-v', '{test_set_input_dir}:/home/{submission_identifier}/te%st',
%  '-w', '/home/{submission_identifier}',
%  '--entrypoint', './{executable_name}',
%  '{docker_image}']

\section{Conclusions}
In this paper, we introduce Twitter's Recsys 2021 Challenge.  This challenge introduces some innovations: the inclusion of a fairness consideration in the evaluation criteria, a novel approach for indirectly incentivizing low-latency solutions, and a requirement that participants respect the changing desires of users-- that tweets that a user has deleted are no longer included in the dataset.  In building this challenge, we balanced several different considerations: fidelity to the real problems at Twitter, user privacy and agency, and  providing data that is of a format and scale that is useful for Recsys practitioners who hope to participate in a Challenge that resembles the type of model-building that takes place at Twitter. We hope this challenge will help shape the direction of similar challenges in the future.  

\section{RecSys 2021 Challenge Summmary}
For RecSys 2021 Challenge, we have two winning groups from Academia and Industry and each group consists of three winning positions. Based on the resources available in Academia and Industry, we have identified three winning solutions from the general leaderboard and the three best solutions whose teams are composed only of academic people. In the following, we provide a brief overview of all the winning solutions. The industry based solutions for twitter user engagement predictions are mainly either ensemble of stacked models or hybrid of various deep neural networks and XGBoost models. The solutions in the order of winning position are:
\begin{itemize}
    \item{1st place:}  \emph{GPU Accelerated Boosted Trees and Deep Neural Networks for Better Recommender Systems}, where authors used an end-to-end GPU-accelerated ensemble of stacked models, using in total 5 XGBoost models and 3 neural networks for engagement predictions.
    \item{2nd place:} \emph{Synerise at RecSys 2021: Twitter user engagement prediction with a fast neural model}, with a simple feed forward neural network that predicts the probability of different engagements types for user and target Tweets by obtaining meaningful Tweet text representation using DistilBERT model and EMD.
    \item{3rd place:} \emph{User Engagement Modeling with Deep Learning and Language Models} where a hybrid Java and Python pipeline was used to extract features from Tweets content which are used to train 4 XGBoost models, one for each engagement type. The authors trained a neural classifier with multi-layer perceptions to predict on users engagement probabilities.
\end{itemize}
We also have three winning position for teams composed only of academic people. The academia solutions are mainly consists of using LightGBM and XGBoost models for Twitter user engagement predictions. The acadmic winners are:

\begin{itemize}
    \item{1st place:}  \emph{Lightweight and Scalable Model for Tweet Engagements Predictions in a Resource-constrained Environment} where the approach used was an optimized LightGBM model which leverages a wide variety of meaningful features from the text of each Tweet and adopted two types of models for engagement prediction: Neural Network (NN) and Gradient Boosting for Decision Tree (GBDT) models.
    \item{2nd place:} \emph{Addressing the cold-start problem with a two-branch architecture for fair tweet recommendation} proposed a two-branch architecture that separates Twitter authors according to their total number of interactions in the dataset. Their solution consists of prediction the user engagement using NN method by identifying similar users and LightGBM models
    \item{3rd place:} \emph{Team JKU-AIWarriors in the ACM Recommender Systems Challenge 2021: Lightweight XGBoost Recommendation Approach Leveraging User Features} used a model that relies on features that can be computed from user engagement counts and exploit a simple XGB classifier, trained on a subset of the training data for twitter user engagement predictions.
\end{itemize}

\section*{Acknowledgements}
The authors would like to thank (in alphabetical order) Robin Burke, Rumman Chowdhury, Michael Ekstrand, Vladislav Grozin and Martha Larson for the fruitful conversations, especially around the topic of fairness. Furthermore the work of Chandler Bair was instrumental in the success of this Challenge.
\bibliographystyle{plain}
\bibliography{biblio}

\end{document}